\documentclass[%
twocolumn,
superscriptaddress,
 amsmath,amssymb,
 aps,
 prr,
]{revtex4-2}
\usepackage{dcolumn}% Align table columns on decimal point
\usepackage{bm}% bold math
\usepackage{graphicx}
\usepackage{color}
\usepackage{comment}
\usepackage[caption=false]{subfig}
\usepackage{color}
\usepackage{bbm}

\begin{document}

\preprint{}

\title{Resetting in Stochastic Optimal Control}% Force line breaks with \\
\author{Benjamin De Bruyne }
\affiliation{LPTMS, CNRS, Univ. Paris-Sud, Universit\'e Paris-Saclay, 91405 Orsay, France}
\author{Francesco Mori}
\affiliation{LPTMS, CNRS, Univ. Paris-Sud, Universit\'e Paris-Saclay, 91405 Orsay, France}

\date{\today}

\begin{abstract}
``When in a difficult situation, it is sometimes better to give up and start all over again''. While this empirical truth has been regularly observed in a wide range of circumstances, quantifying the effectiveness of such a heuristic strategy remains an open challenge. In this paper, we combine the notions of optimal control and stochastic resetting to address this problem. The emerging analytical framework allows not only to measure the performance of a given restarting policy but also to obtain the optimal strategy for a wide class of dynamical systems. We apply our technique to a system with a final reward and show that the reward value must be larger than a critical threshold for resetting to be effective. Our approach, analogous to the celebrated Hamilton-Jacobi-Bellman paradigm, provides the basis for the investigation of realistic restarting strategies across disciplines. As an application, we show that the framework can be applied to an epidemic model to predict the optimal lockdown policy.
\end{abstract}


\maketitle

Finding the optimal strategy to operate a complex system is a longstanding problem and has attracted a lot of attention in the last decades. Since the seminal works of Pontryagin \cite{P87} and Bellman \cite{BK65}, optimal control theory has received renewed interest due to its applications in a wide range of contexts, such as artificial intelligence \cite{RN} and finance \cite{P09}. In a typical setting, optimal control considers a system whose state at time $t$ can be represented by a $d$-dimensional vector $\bm{x}(t)$. For instance, the state $\bm{x}(t)$ could correspond to the degrees of freedom of an autonomous robot or the asset values in a financial portfolio. The system typically evolves in time following a deterministic law, e.g., the laws of motion for mechanical systems or the law of supply and demand for financial markets. Oftentimes, the mathematical modeling of these laws is prohibitively expensive and one introduces a stochastic contribution to account for the missing information on the environment. Given the laws of motion, optimal control aims at operating the system in the best possible way by using an external control, e.g., actuators for robots or market orders in finance.

One of the simplest ways to describe analytically the evolution in time of the system $\bm{x}(t)$ is a first-order differential equation of the form $\dot{\bm{x}}(t)=\bm{f}(\bm{x},t)$. This law is often a simplified description of the system and a source of Gaussian white noise $\bm{\eta}(t)$ is introduced to capture the fluctuations around the deterministic model. In addition, the external control on the system is usually modeled as a drift $\bm{u}(\bm{x},t)$. Summing up these contributions, the full mathematical description of the system is given by
\begin{equation}
\dot{\bm{x}}(t)=\bm{f}(\bm{x},t)+\sqrt{2 D}~\bm{\eta}(t)+\bm{u}(\bm{x},t)\,,
\label{eq:langevin}
\end{equation}
where $\sqrt{2D}$ is the strength of the noise. The external control $\bm{u}(\bm{x},t)$ can be tuned to achieve a given goal, e.g., performing a task for a robot or generating profits in finance. Of course, controlling the system will generate operating costs, such as electrical consumption or transaction fees. Optimally controlling the system requires balancing a trade-off between high rewards, measured over time by a function $R(\bm{x},t)$, and low operating costs, often taken to be proportional to $\bm{u}^2(\bm{x},t)$. To be precise, for a system located at position $\bm{x}$ at time $t$, the reward in a small time interval $dt$ is $R(\bm{x},t)dt$ and the cost is $\bm{u}^2(\bm{x},t)dt/2$.

In principle, solving this optimization problem is intrinsically difficult due to the high dimensionality of the space of solutions. Remarkably, Bellman introduced a general way to solve this problem, known as dynamical programming, which consists in breaking down the optimization into simpler subproblems in a recursive manner such that the present action is taken to maximize the future outcome. In doing so, the key quantity to keep track of is the optimal payoff $J(\bm{x}, t)$, defined as the expected payoff for an optimally controlled system located at ${\bm x}$ at time $t$. Using this framework, one can show that the optimal control is simply given by $\bm{u^*}(\bm{x},t)=\nabla_{\bm{x}} J(\bm{x},t)$, driving the system towards the direction in which the payoff increases the most. The optimal payoff $J(\bm{x},t)$ satisfies the celebrated Hamilton-Jacobi-Bellman (HJB) equation \cite{S93}
\begin{equation}
-\partial_t J =D \Delta_{\bm{x}} J+\bm{f}\cdot\nabla_{\bm{x}} J+\frac12\left(\nabla_{\bm{x}} J\right)^2+R\,,
\label{HJB}
\end{equation}
where $\Delta_{\bm{x}}$ and $\nabla_{\bm{x}}$ are  respectively the Laplacian and the gradient operators. Here for convenience, we dropped the $\bm x$ and $t$ dependence in the functions in equation (\ref{HJB}). The quadratic term $(\nabla_{\bm{x}} J)^2$ renders this equation nonlinear and difficult to solve for arbitrary reward functions. Nevertheless, there exist few analytically solvable cases. For instance, in the case of $d\!=\!1$, where $f(x,t)\!=\!0$ and $R(x,t)\!=\!-\alpha(x-x_f)^2\delta(t-t_f)/2$, the optimal control has the simple form $u^*(x,t)\!=\!-\alpha(x-x_f)/[1+\alpha(t_f-t)]$, which, in the limit $\alpha\to\infty$, is reminiscent of the effective force to generate bridge Brownian motion \cite{MO15}. This optimal control continuously drives the system to maximize the final reward by arriving close to the target $x_f$ at time $t_f$. In more realistic systems, one has to rely on numerical methods to solve Eq.~\eqref{HJB} \cite{Kappen05}.

Ideas from optimal control have also been proven successful in different areas of physics \cite{SS07,BSG15,BSS21}. Moreover, stochastic optimal control has been applied to a variety of systems, such as supply-chain planning \cite{DIS19}, swarms of drones \cite{GTS16}, and fluctuating interfaces \cite{KNO10}. These systems all have in common that the optimal control can be orchestrated as a coordinated sequence of infinitesimal local changes. However, numerous systems do not fall in this class and require global changes to be optimally managed. A timely example of such circumstances is the COVID-19 crisis, during which the main control policies have been global measures such as regional and national lockdowns. Other instances arise in the contexts of search processes, both in the cases of computer algorithms \cite{MZ02} and time-sensitive rescue missions \cite{S06}. In the latter situations, a common and widely observed strategy is to stop and resume the operations from a predefined location. Such situations particularly arise in the contexts of search processes \cite{EM11,EM11b}, chemical reactions \cite{RUK14}, and catastrophes in population dynamics \cite{EF03,VAM10}. Unfortunately, the HJB framework is not well suited to study such resetting protocols. Indeed, resetting is known to be quite different from a local force and exhibits interesting features, including out-of-equilibrium dynamics \cite{MC16,EM16,MMS21}, dynamical phase transitions \cite{KMS14,MSS15,CM15}, and nontrivial first-passage properties \cite{EM11,EM11b} (for a recent review, see \cite{EMS20}). This observation naturally called into question the existence of an analytical framework to devise the optimal resetting control policy.

In this paper, we combine the notions of stochastic resetting and optimal control into \emph{resetting optimal control}, which provides a natural framework to operate systems through stochastic restarts. Our goal  is not to provide an
accurate description of a specific system, but rather
to consider a minimal model to explore resetting in optimal control. To model resetting policies, we exchange the control force $\bm{u}(\bm{x},t)$ for a resetting rate $r(\bm{x},t)$. In a small time interval $dt$, the state of the system is reset to a predefined location $\bm{x}_{\text{res}}$ with probability $p=r({\bm x},t)dt$ and evolves freely otherwise. In sum, the dynamical system evolves according to
\begin{equation}
  \! {\bm x}(t\!+\!dt) \!= \! \left\{\begin{array}{ll}
  \!\!    \bm{x}_{\text{res}}\,,  & \text{prob.}\!=\!p\,,\\
   \!\!{\bm x}(t) \!+\!\bm{f}(\bm{x},t)dt\!+\!\sqrt{2D}{\bm\eta}dt, \!\!\!\!  &\text{prob.}\!=\!1\!\!-\!p\,,
  \end{array}\right.\label{eq:eom}
\end{equation}
where the subscript ``res'' in $\bm{x}_{\text{res}}$ stands for ``resetting location''. Similarly to the HJB framework, we aim at finding the optimal resetting rate $r$, as a function of $\bm{x}$ and $t$, that balances the trade-off between high rewards, measured over time by the function $R(\bm{x},t)$, and low operating costs,  which depend on $r(\bm{x},t)$. To mathematically pose the optimization problem, we naturally extend the HJB paradigm and define the following payoff functional
\begin{equation}
\label{eq:F_resetting}
\!\!\!\mathcal{F}_{\bm{x_0},t}\left[r\right]\!=\!\left\langle\int_{t}^{t_f}\!\!d\tau\left[ R\left(\bm{x}(\tau),\tau\right)\!-\!c\!\left(\bm{x}(\tau),\tau\right)r\!\left(\bm{x}(\tau),\tau\right)\right]\right\rangle_{\!\bm{x_0}}\!\!,
\end{equation}
where $c(\bm{x},\tau)$ is the cost associated to resetting, $t_f$ is the time horizon up to which the system is controlled, and the symbol $\langle\ldots \rangle_{\bm{x_0}}$ indicates the average over all stochastic trajectories starting from $\bm{x_0}$ at time $t$ and evolving according to Eq.~\eqref{eq:eom}. Note that the payoff $\mathcal{F}$ is a functional and depends on the full function $r$.

Remarkably, we find that the optimal resetting policy $r^*$ that maximizes $\mathcal{F}$ is \emph{bang-bang} resulting in an impulsive control strategy \cite{S93}: the system is reset with probability one if its state is outside of a time-dependent domain $\Omega(t)$ and evolves freely otherwise
\begin{equation}
  r^*\left({\bm x},t\right) dt =  \begin{cases}
  0 \,\,\,\,\,\,\text{ if } &\bm{x}\in \Omega(t)\,,\\
  1\,\,\,\,\,\,\text{ if } &\bm{x}\notin \Omega(t)\,.\\  
    \end{cases}\label{eq:rs}
\end{equation}
The domain $\Omega(t)$ evolves according to 
\begin{equation}
  \Omega(t) = \{{\bm x} : J({\bm x},t)\geq J(\bm{x}_{\text{res}},t) -c({\bm x},t) \}\,,\label{eq:Omegad}
\end{equation}
where the optimal payoff function $J(\bm{x},t)=\max_{r}\mathcal{F}_{\bm{x},t}\left[r\right]$ is the solution of the differential equation 
\begin{equation}
-\partial_t J=D \Delta_{\bm{x}} J+\bm{f}\cdot\nabla_{\bm{x}}J+R\,,\,\,\,\,\,\,\bm{x}\in \Omega(t)\,.
\label{HJB_resetting}
\end{equation}
Eq.~\eqref{HJB_resetting} must be solved jointly with Eq.~\eqref{eq:Omegad} starting from the final condition $J(\bm{x},t_f)=0$ and evolving backward in time with the Neumann boundary condition $\nabla_{\bm x} J({\bm x},t) \cdot {\bm n}({\bm x}) = 0$, where ${\bm n}({\bm x}) $ is the normal unit vector to the boundary. Outside of the domain $\Omega(t)$, the solution is given by $J({\bm x},t)= J(\bm{x}_{\text{res}},t) -c({\bm x},t)$. The definition of the domain $\Omega(t)$ in Eq.~\eqref{eq:Omegad} has a clear interpretation: at any given time the optimal policy is to restart the system if its expected payoff is less than the one at the resetting location minus the cost incurred for a restart. The emerging framework outlined in Eqs.~\eqref{eq:rs}, \eqref{eq:Omegad}, and \eqref{HJB_resetting} is the main result of this paper and provides a general method to optimally control stochastic systems through restarts. The derivation of this result is presented in Appendix \ref{app:derivation67}.

\begin{figure}
\centering
\includegraphics[scale=0.5]{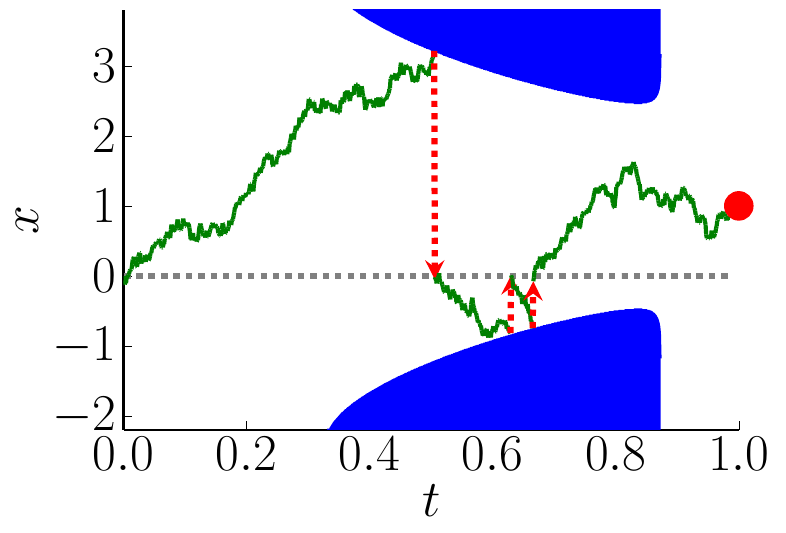}
\caption{Space-time illustration of the optimal resetting policy for a one-dimensional random walk (green line) to reach the target location $x_f=1$ (filled circle) exactly at time $t_f=1$. A reward $\alpha$ is received upon reaching the location $x_f$ at the final time while a unit cost is incurred upon resetting (red dashed arrows) to the origin $x_{\text{res}}=0$ (grey dashed line). The optimal strategy that maximizes the expected payoff is to reset upon touching the blue region and to evolve freely in the white region, which we denote $\Omega(t)$ in the text. The domain $\Omega(t)$ is obtained by numerical integration of the Stefan problem in Eqs.~\eqref{eq:Omegad} and \eqref{HJB_resetting}, with $R(x,t)=\alpha\delta(t-1)\delta(x-1)$, $f(x,t)\!=\!0$, $\alpha=10$ and $D\!=\!1$.  The boundary of the domain $\Omega(t)$ guides the particle to the location $x_f$ at time $t_f$ while avoiding resetting as much as possible. The shape of $\Omega(t)$ depends nontrivially on the reward value $\alpha$. Further explanations for this shape are given in the text.
\label{fig:delta}}
\end{figure}

Going from Eq.~\eqref{eq:F_resetting} to Eq.~\eqref{HJB_resetting}, we have reduced a functional optimization problem to a partial differential equation, which is often easier to solve. Note however that the mathematical problem in Eqs.~\eqref{eq:Omegad} and \eqref{HJB_resetting} is of a special kind as the evolution of the domain of definition $\Omega(t)$ is coupled to the solution $J(\bm{x},t)$ of the differential equation. This kind of equations belongs to a class known as Stefan problems \cite{Crankbook}, which often arise in the field of heat transfer, where one studies the evolution of an interface between two phases, e.g., ice and water on a freezing lake. In this context, one must solve the heat equation for the temperature profile with an interface between the water and ice phases, which moves according to the temperature gradient. The interface is therefore to be considered as an additional unknown function, which must be jointly solved with the differential equation. To draw an analogy with our case, the optimal payoff function $J(\bm{x},t)$ plays the role of the temperature and the boundary of the domain $\Omega(t)$ corresponds to the water-ice interface. Note however that the two Stefan problems have different boundary conditions. The domain $\Omega(t)$ can be obtained by solving the Stefan problem in Eqs.~\eqref{eq:Omegad} and \eqref{HJB_resetting} numerically. This can be achieved, for instance, by using an Euler explicit scheme with a space-time discretization and updating the domain $\Omega(t)$ at each time step according to Eq.~\eqref{eq:Omegad}. The domain $\Omega(t)$ is illustrated in Fig.~\ref{fig:delta} for the case of a one-dimensional random walk with a final reward. Such a situation corresponds to rewarding the system by a finite amount for arriving at some target location at the final time while penalizing it with a unit cost for each resetting event. This setting is further discussed in the next paragraph, where we illustrate our framework with various examples.

\begin{figure}
\centering
\includegraphics[scale=0.5]{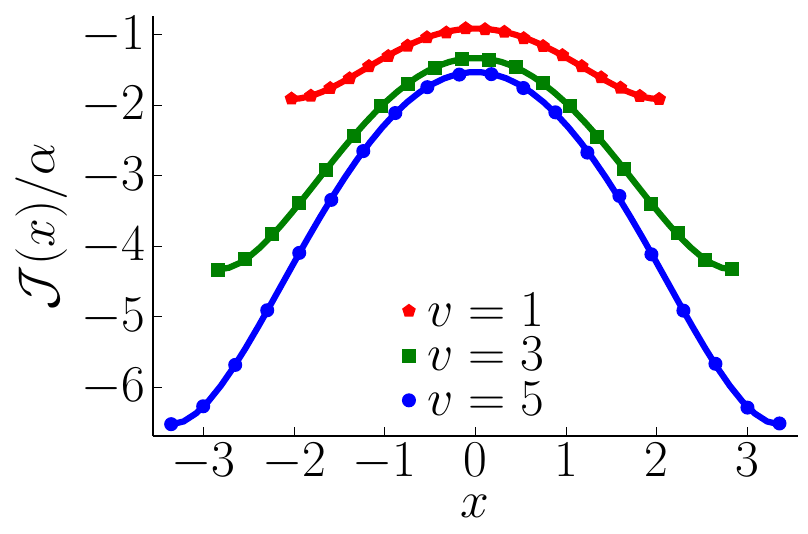} 

\caption{The optimal expected payoff $\mathcal{J}(x)$ for a one-dimensional random walk starting at $x$ for the discounted reward $\mathcal{R}(x)=-\alpha x^2 $ and cost $\mathcal{C}(x)=c$ and for different values of the cost-reward ratio $v\!=\!c/\alpha$. The continuous lines correspond to the exact result in Eq.~\eqref{eq:gyv} while the symbols correspond to numerical simulations performed with $\beta=D=1$. The optimal strategy is to reset for $|x|>u(v)$, where $u(v)$ is defined in the text. For $|x|<u(v)$, the system is let free to evolve and its payoff depends continuously on $x$.
\label{fig:gyv}}
\end{figure}

The Stefan problem in Eq.~\eqref{HJB_resetting} is the analog of the HJB equation \eqref{HJB} for a resetting control. Despite the moving boundary, Eqs.~\eqref{eq:Omegad} and \eqref{HJB_resetting} have a linear dependence in $J$. One might therefore wonder if exactly solvable models exist within this framework. Interestingly, we have found a time-independent formulation in the infinite time horizon limit $t_f\to\infty$ which allows for exact analytical solutions to be found. This is achieved by considering discounted rewards and costs of the form $R\left({\bm x},t\right) = e^{-\beta t}\mathcal{R}\left({\bm x}\right)$ and $c\left({\bm x},t\right) = e^{-\beta t}\mathcal{C}\left({\bm x}\right)$, where $\beta>0$ is the discount rate. Accordingly, we also consider the drift to be time-independent $\bm{f}(\bm{x},t)=\bm{f}(\bm{x})$. Discounted payoffs are common in the control theory literature \cite{FS06} and capture situations in which the effect of the payoff decays over a typical timescale $1/\beta$. Such effect is for instance observed in financial contexts, where $\beta$ is related to interest rates and is used to compare future and present benefits. Using the ansatz $J\left({\bm x},t\right) = e^{-\beta t}\mathcal{J}\left({\bm x}\right)$, we find that Eq.~\eqref{HJB_resetting} becomes a time-independent ordinary differential equation of the form
\begin{equation}
 \beta \mathcal{J}= D \Delta_{\bm{x}}\mathcal{J} +\bm{f}\cdot\nabla_{\bm{x}}\mathcal{J} +\mathcal{R}\,,\quad {\bm x}\in \Omega\,,\label{eq:Jdiffd}
\end{equation}
where the domain $\Omega = \{{\bm x} : \mathcal{J}({\bm x})\geq \mathcal{J}(\bm{x}_{\text{res}}) -\mathcal{C}({\bm x}) \}$ is also independent of time. This equation can be explicitly solved in the absence of external forces by choosing a quadratic reward $\mathcal{R}(x)=-\alpha x^2$ and a constant resetting cost $\mathcal{C}(x)=c$ to the origin $x_{\text{res}}=0$. Note that $\mathcal{R}(x)\leq0$ and is maximized at $x = 0$, rewarding the system for being close to the origin. Solving Eq.~\eqref{eq:Jdiffd}, we obtain, for $\beta\!=\!D\!=\!1$, the exact expression for the optimal payoff
\begin{equation}
\mathcal{J}(x) =\alpha\left[-2-x^2 +2u(v) \frac{\cosh(x)}{\sinh(u(v))}\right]
\,,\quad x\in \Omega\,,
\label{eq:gyv}
\end{equation}
where $u(v)$ is the boundary of the symmetric domain $\Omega$, i.e.,
\begin{equation}
\Omega=\left\{x: \,|x|<u\!\left(v\right)\right\}\,.
\end{equation}
The boundary $u(v)$ is the unique positive solution of the transcendental equation $v-u^2(v) +2u(v)\tanh \left(u(v)/2\right)=0$, where $v=c/\alpha$ is the cost-reward ratio. The optimal strategy thus corresponds to resetting the system if $|x|>u(v)$. When $v\ll 1$, the cost of resetting is much smaller than the reward, therefore the boundary is close to the origin $u(v) \sim \sqrt{2}\,(3v)^{1/4}$, allowing the state of the system to remain close to the optimal location $x=0$. On the other hand, when $v\gg 1$, the cost of resetting is much larger than the running cost and the boundary is set far away from the origin $  u(v) \sim \sqrt{v}+1$. Beyond the optimal resetting policy, our approach predicts the function $\mathcal{J}(x)$, measuring the expected payoff upon starting from $x$ and following the optimal strategy. In Fig.~\ref{fig:gyv}, $\mathcal{J}(x)$ is shown for $|x|<u(v)$ and for various values of the cost-reward ratio $v$. As a function of $x$, $\mathcal{J}(x)$ has a symmetric bell shape centered around the origin, where the reward is maximal. As $|x|$ increases, $\mathcal{J}(x)$ decreases since the reward decreases and the resetting boundary comes closer. Note that, in this special case, the optimal policy can also be recovered within the framework of first-passage resetting \cite{DRR20,DRR21}, as shown in Appendix \ref{app:first}. This is only true for this particular example, where the control variable is a scalar.

Our previous example focused on the case of an infinite time horizon, corresponding to $t_f\to \infty$. We now investigate the effect of a finite time horizon. One of the simplest settings where such effect can be studied is the case of a one-dimensional random walk with a Dirac delta final reward $R(x,t)=\alpha\,\delta(x-x_f)\delta(t-t_f)$, with $\alpha>0$, and a constant resetting cost $c(x,t)=c$. Such parameters correspond to rewarding the system by a finite amount $\alpha$ for arriving at the target position $x_f$ at time $t_f$, while penalizing it with a constant cost $c$ for each resetting event. Note that the system needs to arrive at the location $x_{f}$ exactly at time $t_f$ to get the reward while earlier visits do not provide any additional benefits (as for a \emph{rendez-vous}, where one needs to be at the right place at the right time). The delta reward function has to be understood as the continuum limit of a Gaussian function whose width tends to zero \cite{footnote}. Before presenting the optimal policy for this problem, it is instructive to consider two limiting cases. For $\alpha\to 0$, one should never reset as the reward is not worth the resetting cost. On the other hand, for $\alpha\to\infty$, the cost of resetting becomes negligible and the optimal strategy is to reset if restarting would bring the system closer to $x_f$, independently of time. Interestingly, we observe that the crossover between these two regimes occurs as a sharp transition at the critical value $\alpha=\alpha_c$, where $\alpha_c=x_f c \sqrt{2\pi e}\approx 4.13273 ~x_f c$, which we predict analytically (see Appendix \ref{app:delta}). For $\alpha<\alpha_c$, the optimal policy is to never reset, corresponding to $\Omega(t)=\mathbb{R}$ for all $0\leq t\leq t_f$. The situation is more subtle for $\alpha>\alpha_c$, where resetting is only favorable in a specific time window before $t_f$. To describe this window, it is convenient to introduce the backward time $\tau=t_f-t$. We find that no boundary is present for $\tau<\tau^*$, where $\tau^*$ is the smallest positive solution of the transcendental equation $\alpha e^{-x_f^2/(4D\tau^*)}=c\sqrt{4\pi D\tau^*}$.
At $\tau=\tau^*$, a boundary appears and one must resort to numerical integration techniques to find the solution for $\tau>\tau^*$ (see Fig.~\ref{fig:delta}). We observe numerically that the boundary evolves with $\tau$, i.e., backward in time, in a non-monotonic way and eventually disappears. This optimal policy can be understood as follows. Close to $t_f$, where $\tau<\tau^*$, it is unlikely for the system to reach the target location $x_f$ from the origin in the remaining time. Thus, it is not convenient to reset. On the other hand, for very early times it is not yet necessary to reset since, after resetting, the system would typically evolve away from the target.

 Our framework can be easily generalized to other resetting dynamics. For instance, in light of the recent experiments of stochastic resetting on colloidal particles \cite{TPS20,BBP20,BFP21}, it is relevant to consider the case where the new state of the system ${\bm x}'$ after a resetting event is drawn at random from some probability distribution $P_{\rm R}({\bm x}'|{\bm x})$, which can eventually depend on the current ${\bm x}$ of the system. In practice it is physically impossible to reset the particles to a fixed location, since resetting is performed by using an optic trap. In this case, our main results in Eqs.~\eqref{eq:rs} and \eqref{HJB_resetting} remain valid, while the definition of the domain $\Omega(t)$ is modified as
\begin{equation}
  \Omega(t) = \{{\bm x} : J({\bm x},t)\geq \int d {\bm x}' ~[J({\bm x}',t)P_{\rm R}({\bm x}'|{\bm x})] -c({\bm x},t) \}\,.\label{eq:Omegad_rand}
\end{equation}
Note that the case of a fixed resetting location ${\bm x}_{\rm res}$ corresponds to $P_{\rm R}({\bm x}'|{\bm x})=\delta({\bm x}'-{\bm x}_{\rm res})$. Similarly, one can extend the framework to describe situations in which a finite amount of time is required for each resetting event.

Finally, we illustrate the generality of our framework by applying it on the problem of finding the optimal lockdown policy to navigate a pandemic. We do not aim to provide an accurate description of a specific pandemic, but rather illustrate the generality of our framework and to gain interesting insights. We consider a population of $N$ individuals and we model the evolution of the epidemic with the SIR model, which is one of the simplest compartmental models in epidemiology \cite{SIR}. Previous works have approached this problem by continuously controlling the infection rate \cite{JSSS}. However, in real situations, one cannot adapt the restrictions in a continuous way and our framework is well suited to describe discontinuous policies, such as lockdowns. In Appendix \ref{app:lock}, we show that our technique can be directly applied to this problem to find the optimal time to impose a lockdown.

In sum, we combined optimal control and stochastic resetting to address the effectiveness of restarting policies. The emerging framework, contained in Eqs.~\eqref{eq:Omegad} and \eqref{HJB_resetting}, provides a unifying paradigm to obtain the optimal resetting strategy for a wide class of dynamical systems. Our method can be generalized to discrete-time systems and quadratic costs associated with resetting (see Appendix \ref{app:quad}). Furthermore, it is a simple exercise to include a continuous control policy in addition to resetting to account for realistic systems. In addition, one would need to explore ways to solve the moving boundary problem in high dimensions, which might require approximation schemes. It would be interesting to investigate extensions to optimal stopping problems and to study cost functions that are first-passage functionals \cite{M05}, for instance where the time horizon is a first-passage time. This would be particularly relevant in the context of search processes.

\acknowledgments We warmly thank F. Aguirre-Lopez, M. R. Evans, S. N. Majumdar, S. Redner, A. Rosso, G. Schehr, E. Trizac, and L. Zadnik for fruitful comments and feedback. This work was partially supported by the Luxembourg National Research Fund (FNR) (App. ID 14548297).

\newpage

\onecolumngrid

\appendix

\section{Derivation of Eqs.~\eqref{eq:Omegad} and \eqref{HJB_resetting}}
\label{app:derivation67}

In this appendix, we derive our main result given in Eqs.~6 and 7 of the main text. This derivation is based on a dynamical programming argument (also known as a backward approach). We consider evolving the system from time $t$ to $t+dt$. According to the equations of motion in Eq.~3 of the main text, the state of the system either (i) evolves from $\bm{x}$ to ${\bm x}+\bm{f}(\bm{x},t)dt+ \sqrt{2D}\,{\bm\eta}(t)dt$ with probability $1-r(\bm{x},t)dt$ or (ii) is reset to position $\bm{x}_{\text{res}}$ with probability $r(\bm{x},t)dt$. In the time interval $dt$, the payoff in Eq.~4 of the main text changes by the amount $R(\bm{x},t)dt-c(\bm{x},t)r(\bm{x},t)dt$. For the subsequent evolution of the process from $t+dt$ to $t_f$, the new initial value is either ${\bm x}+\bm{f}(\bm{x},t)dt+ \sqrt{2D}\,{\bm\eta}(t)dt$ in the former case or $\bm{x}_{\text{res}}$ in latter case. Following this argument and using the notation $\mathcal{F}\left[r\,|\,\bm{x},t\right]\equiv \mathcal{F}_{\bm{x},t}\left[r\right]$, we obtain 
\begin{eqnarray}
  \label{eq:Fdt}  \mathcal{F}\left[r\,|\,\bm{x},t\right]&=& R(\bm{x},t)dt-c(\bm{x},t)r(\bm{x},t)dt
  +\left\langle \mathcal{F}\left[r\,\big|\,\bm{x}+\bm{f}(\bm{x},t)dt+ \sqrt{2D}\,{\bm\eta}(t)dt,t+dt\right]\right\rangle (1-r(\bm{x},t)dt)\\&+&\mathcal{F}\left[r\,|\,\bm{x}_{\text{res}},t+dt\right]r(\bm{x},t)dt\,,\nonumber
\end{eqnarray}
where $\langle\ldots\rangle$ is an average over the noise realizations $\bm{\eta}(t)$. The goal is now to maximize both sides over the full function $r$. Note that the functional $\mathcal{F}$ on the right-hand side of Eq.~\ref{eq:Fdt} does not depend on the value of the function $r(\bm{x},t)$ at time $t$, given that its initial time is $t+dt$. We can therefore first maximize over the function $r(\bm{x},\tau)$ for $\tau\in [t+dt,t_f]$ and then over the value $r(\bm{x},t)$ at time $t$, which gives
\begin{eqnarray}
-\partial_t J(\bm{x},t)\!=\!R(\bm{x},t)+D \Delta_x J(\bm{x},t)+\bm{f}(\bm{x},t)\cdot\nabla_{\bm{x}}J(\bm{x},t)+\max_{r(\bm{x},t)}\left\{r(\bm{x},t)\left[J(\bm{x}_{\text{res}},t)-c(\bm{x},t) -J(\bm{x},t) \right]\right\}\,,
\label{eq:Jmax}
\end{eqnarray}
where we used the definition of $J(\bm{x},t)$, expanded to first order in $dt$, and averaged over $\bm{\eta}(t)$. We are left with maximizing a linear function of the scalar variable $r(\bm{x},t)$. Given that $r(\bm{x},t)$ is a Poissonian resetting rate, it must be positive and less than $1/dt$, since $r(\bm{x},t)dt$ is a probability. This immediately gives the expression for the bang-bang optimal control policy $r^*(\bm{x},t)$ given in Eq.~5 in the main text along with the domain $\Omega(t)$. Plugging this back into Eq.~\ref{eq:Jmax}, we obtain the moving boundary problem introduced in the main text.

The Neumann boundary condition $\nabla_{\bm{x}}J(\bm{x},t)\cdot \bm{n}(\bm{x})=0$ can be obtained upon a careful analysis of Eq.~\ref{eq:Fdt} close to the boundary. We start from Eq.~\ref{eq:Fdt} and write
\begin{eqnarray}
\label{eq:F_bound}
 \mathcal{F}\left[r\,|\,\bm{x},t\right]&=& R(\bm{x},t)dt-c(\bm{x},t)r(\bm{x},t)dt
  +\left\langle \mathcal{F}\left[r\,\big|\,\bm{x}+\bm{f}(\bm{x},t)dt+ \sqrt{2D}\,{\bm\eta}(t)dt,t+dt\right]\right\rangle(1-r(\bm{x},t)dt)\nonumber \\& +&\mathcal{F}\left[r\,|\,\bm{x}_{\text{res}},t+dt\right]r(\bm{x},t)dt\,.
\end{eqnarray}
We set $\bm{x}\in \partial \Omega(t)$ and we expand Eq.~\ref{eq:F_bound} to the first nontrivial order in $dt$. The main difference with the case $\bm{x}\in \Omega(t)$, discussed above, is that the linear term in $\bm{\eta}(t)$ does not vanish as the gradient of the payoff functional with respect to $\bm{x}$ vanishes outside of the domain $\Omega(t)$. We obtain
\begin{eqnarray}
0=\sqrt{2D} \nabla_{\bm{x}}\mathcal{F}\left[r\,\big|\,\bm{x},t\right]\cdot \left\langle {\bm\eta}(t)dt~ \mathbbm{1}\left[\bm{x}+ \sqrt{2D}\,{\bm\eta}(t)dt\in \Omega(t)\right]\right\rangle\,,
\label{eq:bound_grad}
\end{eqnarray}
where $\mathbbm{1}$ is the indicator function and the average over $\bm{\eta}(t)$ is of order $O(\sqrt{dt})$. By decomposing $\bm{\eta}(t)$ over the parallel and perpendicular components of the normal direction $\bm{n}(\bm{x})$ to the boundary at $\bm{x}$, Eq.~\ref{eq:bound_grad} immediately gives the boundary condition $\nabla_{\bm{x}}\mathcal{F}\left[r|\bm{x},t\right]\cdot\bm{n}(\bm{x})=0$, which is by definition also valid for $J$.

\section{First-passage resetting in a finite interval}
\label{app:first}

\begin{figure}
\includegraphics[scale=0.5]{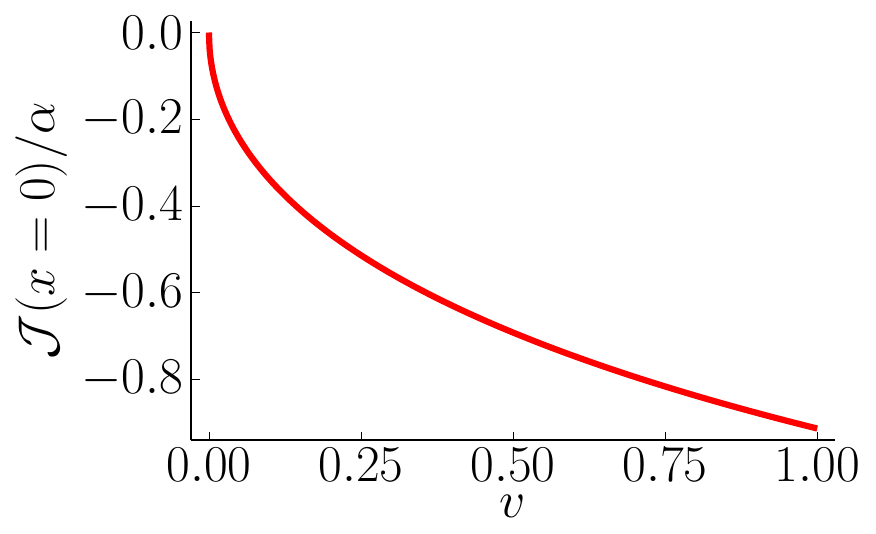}
\caption{The rescaled optimal expected payoff $\mathcal{J}(x)$ for a one-dimensional random walk starting at $x=0$ for the discounted reward $\mathcal{R}(x)=-\alpha x^2 $ and cost $\mathcal{C}(x)=c$ as a function of the cost-reward ratio $v\!=\!c/\alpha$. \label{fig:Ja}}
\end{figure}

In this appendix, we show that the optimal policy in the special case of discounted payoffs can be obtained within the framework of first-passage resetting. We consider a one-dimensional diffusive particle with diffusion coefficient $D$ whose position $x$ belongs to the finite interval $[-L,L]$ with $L>0$. We assume that when the particle reaches any of the interval boundaries it is instantaneously reset to the origin $x_{\text{res}}=0$. We assume that at the initial time the particle is at position $x_0$, with $-L<x_0<L$. In analogy with the discounted payoff in the main text, we associate a discounted reward $R(x,t)= e^{-\beta t}\mathcal{R}\left(x\right)$ and discounted cost of resetting $c(x,t)=e^{-\beta t}\mathcal{C}\left(x\right)$ to the trajectory of the particle. Considering an infinite time-horizon, the average payoff is therefore given by
\begin{eqnarray}
\mathcal{J}(x|L)=\left\langle \int_{0}^{\infty}d\tau ~ e^{-\beta \tau}\mathcal{R}\left(x\right)-\sum_{i=1}^{\infty}e^{-\beta t_i}\mathcal{C}\left(x\right)\right\rangle_{x_0}\,,\label{eq:JxL}
\end{eqnarray}
where $t_i$ is the time of the $i$-th resetting event and the average is computed over all trajectories $\{x(\tau)\}$ with $x(0)=x_0$. In particular, for $\mathcal{R}(x)=-\alpha x^2$ and $\mathcal{C}(x)=c$, we obtain
\begin{eqnarray}
\mathcal{J}(x|L)=-\alpha \int_{0}^{\infty}d\tau ~ e^{-\beta \tau }\left\langle x(\tau)^2\right\rangle_{x_0}-c\sum_{i=1}^{\infty} \left\langle e^{-\beta t_i}\right\rangle_{x_0} \,.\label{eq:JxL2}
\end{eqnarray}
The two average quantities can be computed by standard renewal techniques and one obtains
\begin{eqnarray}
\mathcal{J}(x|L)=-\alpha \left[\frac{2D}{\beta^2}+\frac{x_0^2}{\beta}-\frac{L^2}{\beta}\frac{\cosh(x_0\sqrt{\beta/D})}{\cosh(L\sqrt{\beta/D})-1}\right]-c\left[\frac{\cosh(x_0\sqrt{\beta/D})}{\cosh(L\sqrt{\beta/D})-1}\right] \,.\label{eq:JxL3}
\end{eqnarray}
Upon minimizing over $L$ and setting $\beta=D=1$, we recover the expression in Eqs.~9 and 10 in the main text. The optimal payoff in Eq.~9 in the main text is shown as a function of the cost-reward ratio in Fig.~\ref{fig:Ja}. The optimal payoff decreases monotonically as the cost-reward ratio is increased.

\section{Dirac delta final reward}

\label{app:delta}

\begin{figure}
\includegraphics[scale=0.5]{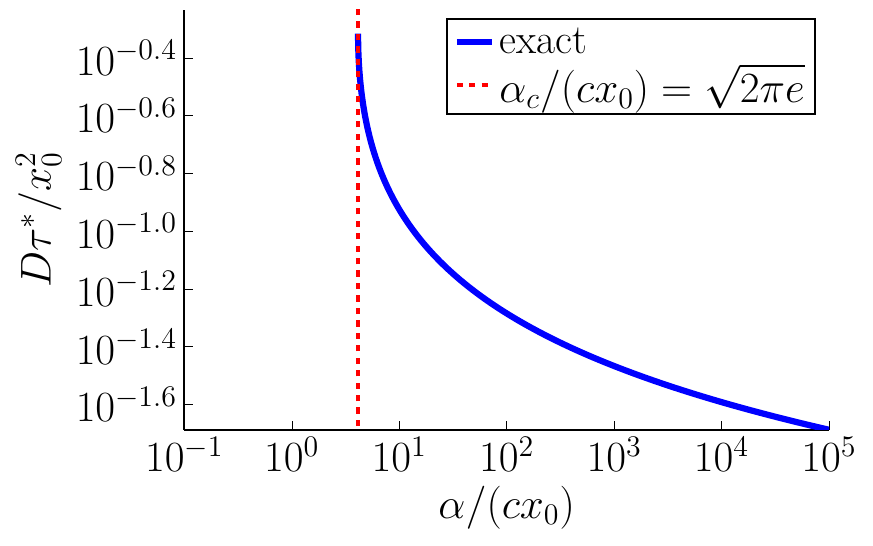}
\caption{Scaled time $D\tau^*/x_0^2$ at which the barrier first appears as a function of $\alpha/(cx_0)$. The continuous blue line corresponds to the smallest positive solution of 
$\alpha=c\sqrt{4\pi D\tau} ~e^{x_f^2/(4D\tau)}$, while the dashed red line corresponds to the critical value $\alpha_c=\sqrt{2\pi e}cx_0$. For $\alpha<\alpha_c$, no barrier is present. \label{fig:tau}}
\end{figure}

In this appendix, we consider the case of a Dirac delta final reward in $d=1$, corresponding to $R(x,t)=\alpha\delta(x-x_f)\delta(t-t_f)$, where $\alpha>0$ is the magnitude of the reward and $x_f$ is the target location. We also assume that $f(x,t)=0$, $c(x,t)=c>0$ and we consider the optimal payoff in the time-reversed dynamics, defined as $I(x,\tau)=J(x,t_f-\tau)$. It is easy to show that $I(x,\tau)$ satisfies the diffusion equation
\begin{eqnarray}
\partial_{\tau}I(x,\tau)=D\partial_{xx}I(x,\tau)\,,
\end{eqnarray}
with initial condition $I(x,\tau=0)=\alpha\delta(x-x_f)$. Assuming that no boundary appears for small $\tau$ (to be verified a posteriori), the optimal payoff function is the usual Gaussian profile
\begin{eqnarray}
I(x,\tau)=\alpha\frac{1}{\sqrt{4\pi D \tau}}e^{-(x-x_f)^2/(4 D\tau)}\,.\label{eq:I_tau}
\end{eqnarray}
A boundary appears at time $\tau$ when the condition $I(x,\tau)<I(0,\tau)-c$ is verified for the first time for some value of $x$. Using Eq.~\ref{eq:I_tau}, this condition can be rewritten, for $|x-x_f|>x_f$, as
\begin{equation}
\alpha>\frac{c\sqrt{4\pi D\tau}}{e^{-x_f^2/(4D\tau)}-e^{-(x-x_f)^2/(4D\tau)}}\,.\label{eq:cond2}
\end{equation}
Minimizing the right-hand side of Eq.~\ref{eq:cond2} over $x$ and $\tau$, we obtain $\alpha>cx_f\sqrt{2e\pi}$. Thus, for $\alpha<\alpha_c=cx_f\sqrt{2e\pi}$, no boundary appears and the cost function is given by Eq.~\ref{eq:I_tau} for any $\tau$. On the other hand, for $\alpha>\alpha_c$, two boundaries appear at time $\tau^*$, which is the smallest solution of
$\alpha=c\sqrt{4\pi D\tau} ~e^{x_f^2/(4D\tau)}$. Thus, for $\tau<\tau^*$, the cost function is given by Eq.~\ref{eq:I_tau}, while it is hard to determine analytically for $\tau>\tau^*$. We obtain numerically the boundary for $\tau>\tau^*$ (see Fig. 1 of the main text). Note that at $\tau=\tau^*$ the condition in Eq.~\ref{eq:cond2} is only verified for $x\to \pm\infty$, meaning that the two boundaries start from infinity at the critical time. The critical time $\tau^*$ is shown in Fig. \eqref{fig:tau} as a function of $\alpha$. The asymptotic behaviors of $\tau^*$ as a function of $\alpha$ are given by
\begin{eqnarray}
\tau^*=\begin{cases}
x_f^2/(2D)\,, \quad&\text{  for   }\alpha\to \alpha_c\,,\\
[x_f^2/(4D)]/\log(\alpha)\,, &\text{  for   }\alpha\to \infty\,.
\end{cases}
\end{eqnarray}

\section{Application to lockdown policies}

\label{app:lock}

In this appendix, we apply the framework developed in the main text to a problem of finding the optimal lockdown policy to navigate a pandemic. To describe the propagation of an epidemic within a population, we employ the SIR model. We consider a population of $N$ individuals, divided into three groups $(S,I,R)$, where $S$, $I$, and $R$ are respectively the number of susceptible, infected, or recovered individuals. Note that by definition $S+I+R=N$ and therefore a given state of the system is completely specified by the vector ${\bm x}=(S,I)$. We assume that new infections occur at a rate $\beta S I$, while infected individuals recover at a rate $\gamma I$. In other words, the system evolves as a Markov jump process with rates
\begin{equation}
w[(S,I)\to (S-1,I+1)] = \beta	S I\,,
\end{equation}
and 
\begin{equation}
w[(S,I)\to (S,I-1)] = \gamma I\,,
\end{equation}
where $w[{\bm x}\to {\bm x'}]$ indicates the rate of the transition from ${\bm x}$ to $ {\bm x'}$.

During a real pandemic, national governments are usually presented with conflicting objectives. Indeed, the rapid spread of the disease carries heavy public-health costs, in particular when the number of infected individuals that require treatment exceeds the hospital alert level of a country. This spread can be countered by imposing lockdowns, which however have an impact on the economy of a country. 

\begin{figure}
\includegraphics[scale=0.5]{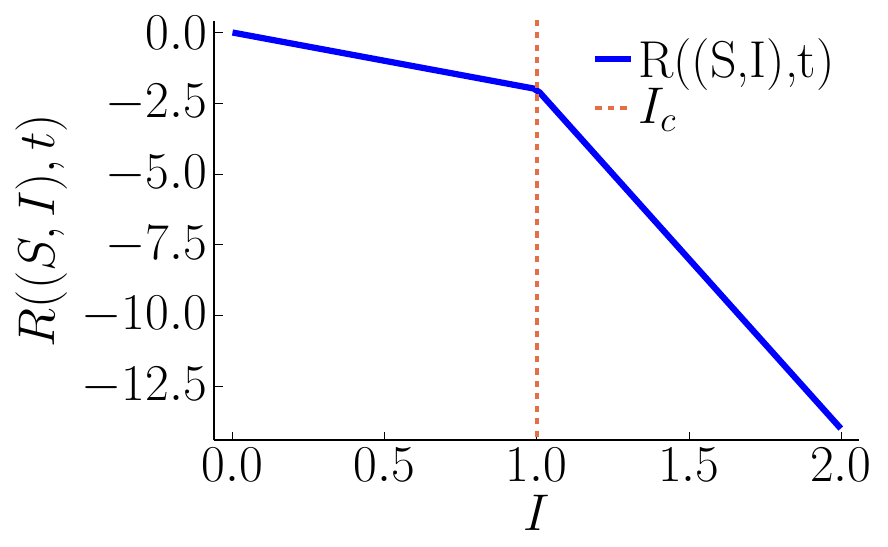}
\caption{Reward function $R((S,I),t)$ in (\ref{eq:Rld}) describing the public-health cost in the context of the optimal lockdown policy problem. The reward is a piece-wise linear function whose slope increases above the hospital alert level threshold $I_c$.  \label{fig:rewardLD}}
\end{figure}

To describe the public-health cost, we introduce the (negative) reward function (see Fig.~\ref{fig:rewardLD})
\begin{equation}
R((S,I),t)=-a I-[b(I-I_c)]~ \theta(I-I_c)\,,\label{eq:Rld}
\end{equation}
where $I_c$ represents the hospital alert level, $a>0$ is the cost per infected person when hospital are not saturated and $b>0$ is the excess cost when the hospitals are saturated. Here $\theta(z)$ is the Heaviside theta function.
 We describe lockdowns by allowing the possibility for governments to decrease the number of infected people by a constant fraction, without changing the number of susceptible people. In other words, the effect of a lockdown is to reset the system from state ${\bm x}=(S,I)$ to state ${\bm x'}=(S, \lfloor \alpha I\rfloor)$, where $0<\alpha<1$. We assume that each lockdown comes with a fixed cost $c>0$, due to the negative socio-economic impact. The goal in then to find the optimal resetting (or lockdown) policy $r^*({\bm x},t)$ which minimizes the payoff function defined in Eq.~(4) of the main text, in order to control the system up to a given time horizon $t_f$.

It is easy to employ our framework in this case and to show that the optimal payoff function $J((S,I),t)$ evolves according to
\begin{equation}
-\partial_t J((S,I),t)=R((S,I),t)+\beta S I \left[J((S-1,I+1),t)-J((S,I),t)\right]-\gamma I\left[J((I-1,S),t)-J((S,I),t)\right]\,,
\label{stefan_supmat}
\end{equation}
for $(S,I)\in \Omega(t)$ and 
\begin{equation}
J((S,I),t)=J((S,	\lfloor \alpha I\rfloor),t)-c\,,
\end{equation}
for $(S,I)\notin \Omega(t)$. Here the domain $\Omega(t)$ is defined as
\begin{equation}
\Omega(t)=\{{\bm x}:J((S,I),t)\geq J((S,	\lfloor \alpha I\rfloor),t)-c\}\,.
\end{equation}

\begin{figure}
\centering
\includegraphics[scale=0.5]{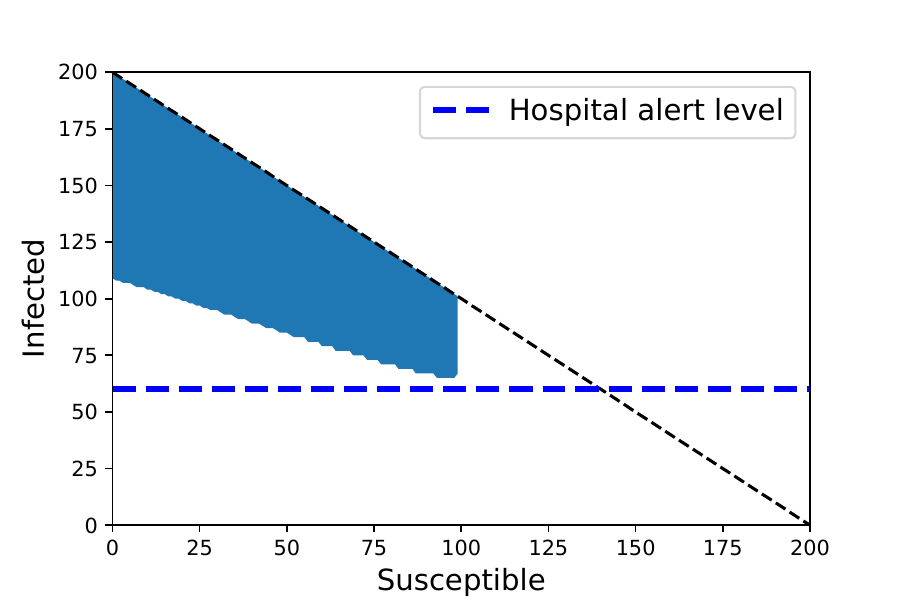}
\caption{Phase-space illustration of the optimal lockdown policy for a SIR epidemic model with infection rate $\beta$ and recovery rate $\gamma$. Since the total number $N$ of individuals is fixed, only configurations below the black dashed line, where $S+I\leq N$, are allowed. We consider a (negative) reward $R((S,I),t)=-a I-[b(I-I_a)]~ \theta(I-I_a)$, where $I_a$ is the hospital alert level (horizontal dashed line) and $\theta(x)$ is the Heaviside step function. A fixed cost $c$ is incurred upon resetting from the state $(S,I)$ to the state $(S, \lfloor\alpha I\rfloor)$, which describes the effects of a lockdown. The optimal strategy that maximizes the expected payoff is to reset upon touching the blue region and to evolve freely in the white region, which we denote $\Omega(t)$ in the text. The domain $\Omega(t)$ is obtained by numerical integration of the Stefan problem presented in \eqref{stefan_supmat}. The boundary of the domain $\Omega(t)$ guides the epidemic below the hospital alert level while avoiding lockdowns as much as possible. The figure corresponds to the choice of parameters $t=0$, $a=2$, $b=10$, $\alpha=0.5$, $N=200$, $I_a=0.3~ N$, $\beta=5/N$, $t_f=1$, and $\gamma=1$. An optimally-controlled epidemic trajectory is shown in \ref{fig:traj}.
\label{fig:SIR}}
\end{figure}

\begin{figure}
\includegraphics[scale=0.5]{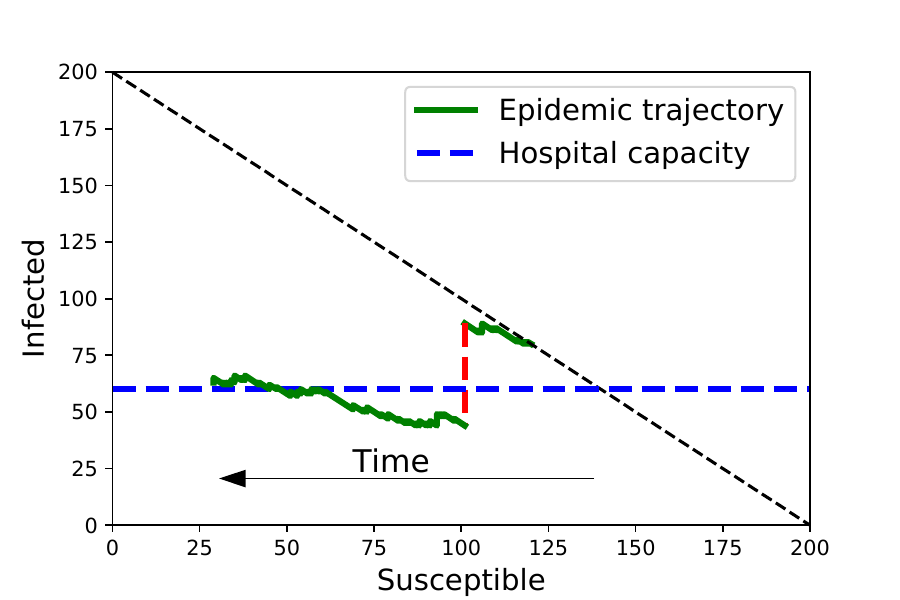}
\caption{Epidemic evolution in the SIR model controlled by an optimal lockdown policy according to the reward function given in (\ref{eq:Rld}) and a fixed cost $c$ for resetting. The red dashed line corresponds to the lockdown event. Since the total number $N$ of individuals is fixed, only configurations below the black dashed line, where $S+I\leq N$, are allowed. The horizontal dashed line describes hospital alert level $I_c$. The figure corresponds to the choice of parameters $a=2$, $b=10$, $\alpha=0.5$, $N=200$, $I_c=0.3~ N$, $\beta=5/N$, and $\gamma=1$, $t_f=1$, which are the same as the one used in Fig.~\ref{fig:SIR}. } \label{fig:traj}

\end{figure}

We employ our framework numerically and obtain the optimal policy in the $(S-I)$ plane as shown in Fig.~\ref{fig:SIR}. As can be seen in this figure, the policy is to never reset if the number of susceptible individuals is larger than a critical value. Then, as the number of susceptible individuals is lowered, the optimal policy is to reset if the number of infected individuals is larger than a threshold which increases as the number of susceptible individuals is lowered. The optimal domain $\Omega(t)$ guides the epidemic below the hospital alert level while avoiding lockdowns as much as possible. An optimally-managed epidemic trajectory is shown in Fig.~\ref{fig:traj}.

\section{Generalization to a quadratic resetting cost}
\label{app:quad}

In this appendix, we generalize our framework to the case in which a quadratic cost is associated to resetting with the following payoff functional
\begin{eqnarray}
   \mathcal{F}[r|{\bm x}_0,t] = \left\langle  \int_{t}^{t_f} d\tau \,\left[R\left({\bm x}(\tau),\tau\right) - \frac{1}{2}c\left({\bm x}(\tau),\tau\right)r\left({\bm x}(\tau),\tau\right)^2 \right]\right\rangle_{{\bm x}_0}\,, \label{eq:G2}
\end{eqnarray}
with $c(\bm{x},\tau)\geq 0$. Following the procedure outlined in the main text, we obtain the moving boundary problem
\begin{eqnarray}
  \begin{array}{rll}
  -\partial_{t} J({\bm x},t) &=  D \Delta_{\bm x}J({\bm x},t) +\bm{f}({\bm x},t)\cdot\nabla_{\bm{x}}J({\bm x},t) + R\left({\bm x},t\right)\,, & {\bm x}\in \Omega(t)\,,\\
 -\partial_{t}J({\bm x},t)&= D \Delta_{\bm x}J({\bm x},t) +\bm{f}({\bm x},t)\cdot\nabla_{\bm{x}}J({\bm x},t) + R\left({\bm x},t\right)+ \frac{1}{2\,c\left[{\bm x}(\tau),\tau\right]}[J({\bm x},t)-J({\bm 0},t)]^2\,,& {\bm x} \notin \Omega(t)\,,
\end{array}\label{eq:Jdiff}
\end{eqnarray}
where 
\begin{eqnarray}
\Omega(t)=\{\bm{x}: J({\bm x},t)\geq J(\bm{x}_{\text{res}},t) \}\,.
 \end{eqnarray}
The differential equation \ref{eq:Jdiff} must by solved by imposing the continuity of the solution and its derivative at the boundary of $\Omega(t)$. The optimal resetting rate $r^*(\bm x,t)$ is no more ``bang-bang'' and is given by
\begin{eqnarray}
  r^*(\bm x,t) = \left\{\begin{array}{ll}
  0\,, & {\bm x} \in \Omega(t)\,,\\ 
  \left[J(\bm{x}_{\text{res}},t)-J({\bm x},t)\right]/c\left[{\bm x}(\tau),\tau\right] \,,& {\bm x} \notin \Omega(t)\,.
  \end{array}\right.\label{eq:rs2}
\end{eqnarray}

\section{Generalization to discrete-time systems}

\label{app:discrete}

In this appendix, we generalize our framework to the case of discrete-time systems whose state ${\bm x}_n$ evolves according to the Markov rule
\begin{eqnarray}
   {\bm x}_n = \left\{\begin{array}{ll}
    \bm{x}_{\text{res}} ,&\quad\text{with probability } p_n({\bm x})\\
    {\bm x}_{n-1} + f_n({\bm x_{n-1}})+ {\bm \eta_n}\,, &\quad\text{with probability }1-p_n({\bm x})\,,
  \end{array}\right.\label{eq:eomd}
\end{eqnarray}
where ${\bm \eta_n}$ are independently and identically distributed random variables drawn from a probability distribution $q({\bm \eta})$ and $f_n({\bm x})$ is an external force. The distribution $q({\bm \eta})$ is arbitrary and includes for instance the case of fat-tailed distributions. The payoff functional in Eq.~4 of the main text straightforwardly generalizes to 
\begin{eqnarray}
   \mathcal{F}(\{p_m,\ldots,p_n\}|{\bm x}_m,m) = \left\langle \sum_{l=m}^n \,R\left({\bm x}_l,l\right) - c({\bm x}_l,l)p_l({\bm x_l}) \right\rangle_{{\bm x}_m}\,, \label{eq:Gd}
\end{eqnarray}
where the control is the sequence of resetting probabilities $\{p_m,\ldots,p_n\}$ and the average on the right-hand side is taken over all system trajectories starting from ${\bm x_m}$ at step $m$. Following a similar approach as in the previous sections, we obtain that the optimal policy is given by
\begin{eqnarray}
  p_m\left({\bm x}\right)  =  \begin{cases}
  0 \,\,\,\,\,\,\text{ if } &\bm{x}\in \Omega_m\,,\\
  1\,\,\,\,\,\,\text{ if } &\bm{x}\notin \Omega_m\,,\\  
    \end{cases}
\end{eqnarray}
where $\Omega_m\subseteq 	\mathbb{R}^d$ is a time-dependent domain. The corresponding discrete moving boundary problem for the optimal payoff function $J(\bm{x},m)=\max_{\{p_m,\ldots,p_n\}}\mathcal{F}\left[\{p_m,\ldots,p_n\}\,|\,\bm{x},m\right]$ is given by
\begin{eqnarray}
  J(\bm{x},m) = R\left({\bm x},m\right) + \max\left[J(\bm{x}_{\text{res}},m+1)+c({\bm x},m), \int_{-\infty}^\infty d{\bm \eta} \,q({\bm \eta})\,J(\bm{x}+f_m({\bm x})+{\bm \eta},m+1) \right]\,,\label{eq:Jdist}
\end{eqnarray}
with
\begin{eqnarray}
  \Omega_m = \left\{\bm{x}: \int_{-\infty}^\infty d{\bm \eta} \,p({\bm \eta})\,J(\bm{x}+f_m({\bm x})+{\bm \eta},m+1) \geq J(\bm{x}_{\text{res}},m+1)+c({\bm x},m)\right\}\,.\label{eq:omegam}
\end{eqnarray}

\setcounter{secnumdepth}{2}

\end{document}